\documentclass[10pt]{article}

\usepackage[a4paper, margin=1in]{geometry} 
\usepackage{amsmath, amssymb, amsfonts}
\usepackage{bm}
\usepackage{graphicx}
\usepackage{palatino}
\usepackage{mathpazo}
\usepackage{textcomp}
\usepackage{float}
\usepackage{booktabs}
\usepackage{dcolumn}
\usepackage{multirow}
\usepackage{hyperref}
\usepackage{xcolor}
\usepackage{orcidlink}

\hypersetup{colorlinks=true, citecolor=blue}
\begin{document}
	
	\title{\textbf{Comparative Study of $f(T)$ Gravity Models with Observational Constraints\\
			from \textit{OHD} and \textit{Pantheon+ datasets}}}
	
	\author{
		Anirudh Pradhan\orcidlink{0000-0002-1932-8431}\\
		\small Centre for Cosmology, Astrophysics and Space Science (CCASS),\\
		\small GLA University, Mathura-281406, U.P., India\\
		\small Email: pradhan.anirudh@gmail.com
		\and
		S. A. Salve \orcidlink{0009-0009-3293-6915}\\
		\small Department of Mathematics, Jijamata Mahavidyalaya,\\
		\small Buldana – 443001, Maharashtra, India\\
		\small Email: salvesanjaya@gmail.com
		\and
		V. B. Raut \orcidlink{0009-0003-3639-9578}\\
		\small Department of Mathematics, Mungsaji Maharaj Mahavidyalaya,\\
		\small Darwha – 445202, Maharashtra, India\\
		\small Email: drvilasraut@gmail.com
		\and
		S. H. Shekh \orcidlink{0000-0003-4545-1975}\\
		\small Department of Mathematics, S.P.M. Science and Gilani Arts, Commerce College,\\
		\small Ghatanji, Yavatmal, Maharashtra 445301, India\\
		\small L. N. Gumilyov Eurasian National University, Kazakhstan\\
		\small Pacif Institute of Cosmology and Selfology (PICS), Odisha, India\\
		\small Email: da\_salim@rediff.com
	}
	
	\date{}
	
	\maketitle
	
	\begin{abstract}
		The late-time acceleration of the universe remains one of the most significant open problems in modern cosmology. Modified gravity frameworks such as $f(T)$ gravity provide a geometric alternative to dark energy by attributing cosmic acceleration to torsional effects. In this study, we present a comparative analysis of three different forms of $f(T)$ models: (i) a simple power-law form $f(T) = \eta (-T)^{n}$, (ii) the exponential form $f(T) = \beta T_{0}\left(1-e^{-p \sqrt{T/T_{0}}}\right)$ and (iii) a logarithmic form $f(T) = \gamma T \ln\!\left(\frac{T}{T_{0}}\right)$. Using parameterization of the deceleration parameter $q(z)$ and the corresponding $H(z)$ expression, we constrain the model parameters with the recent Hubble parameter and BAO data through a Markov Chain Monte Carlo (MCMC) approach. The physical behavior of the effective energy density, equation of state parameter, squared sound speed, cosmological $Om(z)$ diagnostics, and energy conditions (NEC, DEC, SEC) were investigated for all three models. Our comparative analysis shows that all models asymptotically approach the $\Lambda$CDM behavior at late times, while they differ in stability properties and energy condition behaviors. In particular, the violation of the strong energy condition (SEC) has emerged as a common feature consistent with current accelerated expansion. This study highlights how different $f(T)$ functional forms can yield distinct cosmological dynamics while maintaining consistency with observational data.
		
		\vspace{2mm}
	\end{abstract}
	
	\section{Introduction}
	
	Over the past two decades, cosmological observations including Type Ia supernovae \cite{riess1998,perlmutter1999}, cosmic microwave background (CMB) anisotropies \cite{spergel2003,spergel2007}, and large-scale structure surveys \cite{tegmark2004,eisenstein2005} have confirmed that the universe is undergoing a phase of accelerated expansion. This unexpected discovery has given rise to the dark energy problem, motivating both new physical components with negative pressure and modifications of the gravitational sector itself. Over the last 25 years, there has been a significant increase in the number of Type Ia supernovae discovered as well as the quality of fundamental data supplied. Distances to Type Ia events have provided statistically correct values for the Hubble constant, $\Omega_M$, and $\Omega_\Lambda$. Among the various approaches, modified theories of gravity offer compelling alternatives to explain late-time cosmic acceleration without invoking an exotic dark energy fluid.
	
		General relativity is a gravitational theory that accurately describes some of gravity's effects, such as solar system tests, gravitational lensing, gravitational waves, black holes, and so on, within a specific framework of homogeneous and isotropic space-time. However, the validity range of general relativity may be limited due to the abundance and nature of dark energy and dark matter, the nature of inflation, the Hubble tension $H_0$ and $S_{8}$ tension, the possible values of local anisotropy in the evolution of the Universe, and the theoretical problems of the cosmological constant and non-renormalizability. Modified gravity theories use a variety of ways to extend the form of general relativity, resulting in new field equations and, hence, cosmological ramifications.  They play an important part in and contribute to modern cosmology, laying the groundwork for our current knowledge of physical events in the Universe. \\
	One of the most actively explored frameworks is $f(T)$ gravity, a generalization of teleparallel gravity (TG), where the torsion scalar $T$ replaces the Ricci scalar $R$ in the action, and the Lagrangian is generalized to an arbitrary function $f(T)$ \cite{bengochea2009,ferraro2007,linder2010,N1,N2,N3,N4}. Unlike $f(R)$ gravity, which leads to fourth-order field equations, $f(T)$ gravity results in second-order equations, which make it mathematically more tractable. Within this framework, cosmic acceleration arises from torsional contributions, which effectively mimic a dark energy component. \\
	Recently, various functional forms of $f(T)$ have been proposed and investigated. The power-law version of $f(T)$ gravity describes the gravitational action as a power-law function of the torsion scalar, $T$. This model is used in cosmology to analyze the expansion of the universe and is restricted by cosmological data such as the Hubble and Pantheon supernovae datasets. The power-law form $f(T) = \eta (-T)^{n}$ has been shown to yield quintessence-like behavior \cite{myrzakulov2011}, while the exponential form  $f(T) = \beta T_{0}\left(1-e^{-p \sqrt{T/T_{0}}}\right)$ produces a smooth evolution consistent with observational data \cite{nassur2015}. A combined power-logarithmic form $f(T) = \gamma T \ln\!\left(\frac{T}{T_{0}}\right)$ has also been suggested to capture richer dynamics \cite{delcampo2012}. Despite these efforts, a systematic comparison of such models under the same parameterization and dataset constraints remains limited. \\
	The present study aims to fill this gap by performing a comparative analysis of three $f(T)$ gravity models using a parameterized deceleration parameter $q(z)$. By employing the corresponding Hubble parameter $H(z)$, we constrained the model parameters using the  updated Hubble parameter and BAO datasets. Subsequently, we investigate the physical and cosmological consequences of each model, including the behavior of the effective energy density, equation of state, squared speed of sound, statefinder and $Om(z)$ diagnostics, and the fulfillment or violation of classical energy conditions. Our results demonstrate that while all three models tend toward $\Lambda$CDM behavior at late times, their differences in stability and energy condition features provide crucial insights for distinguishing between alternative $f(T)$ scenarios.\\
	The structure of this article is as follows: In Section \ref{II}, we provide a brief overview of the $f(T)$ gravity framework. In Sections \ref{III} and \ref{IV} we introduced the parameterization of $Om(z)$ and the observational datasets used for constraints. Section \ref{V} presents comparative results of the different $f(T)$ models.  Finally, Section \ref{VI} summarizes the conclusions and provides comparative insights of this study.
	
	\section{Formulation of $f(T)$ Gravity}\label{II}
	Teleparallel gravity offers an alternative geometrical description of gravitation, in which torsion rather than curvature plays a central role. In this framework the dynamical variables are the tetrad (vierbein) fields $e^{A}_{\ \mu}$, which form an orthonormal basis for the tangent space at each space-time point. The spacetime metric is then related to the tetrads through
	\begin{equation}\label{e1}
		g_{\mu\nu} = \eta_{AB}\, e^{A}_{\ \mu} e^{B}_{\ \nu},
	\end{equation}
	where $\eta_{AB} = \text{diag}(-1,1,1,1)$ is the Minkowski metric. The determinant of the tetrad field is denoted as $e = \text{det}(e^{A}_{\ \mu}) = \sqrt{-g}$.
	
	In teleparallel geometry the gravitational interaction is described by the Weitzenböck connection,
	\begin{equation}\label{e2}
		\Gamma^{\lambda}_{\ \mu\nu} = e^{\lambda}_{A}\, \partial_{\nu} e^{A}_{\ \mu},
	\end{equation}
	which is curvature-free but possesses torsion. The corresponding torsion tensor is defined as follows:
	\begin{equation}\label{e3}
		T^{\lambda}_{\ \mu\nu} = \Gamma^{\lambda}_{\ \nu\mu} - \Gamma^{\lambda}_{\ \mu\nu}
		= e^{\lambda}_{A}\left(\partial_{\mu} e^{A}_{\ \nu} - \partial_{\nu} e^{A}_{\ \mu}\right).
	\end{equation}
	
	From the torsion tensor one constructs the contorsion tensor
	\begin{equation}\label{e4}
		K^{\mu\nu}_{\ \ \lambda} = -\tfrac{1}{2}\left(T^{\mu\nu}_{\ \ \lambda} - T^{\nu\mu}_{\ \ \lambda} - T_{\lambda}^{\ \mu\nu}\right),
	\end{equation}
	and the superpotential
	\begin{equation}\label{e5}
		S_{\lambda}^{\ \mu\nu} = \tfrac{1}{2}\left(K^{\mu\nu}_{\ \ \lambda} + \delta^{\mu}_{\lambda} T^{\alpha\nu}_{\ \ \alpha} - \delta^{\nu}_{\lambda} T^{\alpha\mu}_{\ \ \alpha}\right).
	\end{equation}
	The torsion scalar is then defined as
	\begin{equation}\label{e6}
		T = S_{\lambda}^{\ \mu\nu} T^{\lambda}_{\ \mu\nu}.
	\end{equation}
	
	The teleparallel equivalent of general relativity (TEGR) is obtained from the action
	\begin{equation}\label{e7}
		S_{\rm TEGR} = \frac{1}{16\pi G}\int d^{4}x\, e\, T + S_{m},
	\end{equation}
	where $S_{m}$ represents the matter action. A natural extension of this theory, known as $f(T)$ gravity, generalizes the torsion scalar $T$ to an arbitrary function $f(T)$, leading to action
	\begin{equation}\label{e8}
		S = \frac{1}{16\pi G}\int d^{4}x\, e\, f(T) + S_{m}.
	\end{equation}
	
	Varying this action with respect to the tetrad fields yields the modified field equations:
		\begin{equation}\label{e9}
			\left[e^{-1}\partial_{\mu}\big(e\, S_{A}^{\ \mu\nu}\big) + e_{A}^{\ \lambda} T^{\rho}_{\ \mu\lambda} S_{\rho}^{\ \nu\mu}\right]f_{T}
			+ S_{A}^{\ \mu\nu}\,\partial_{\mu}(T)\, f_{TT}
			+ \tfrac{1}{4}e_{A}^{\ \nu} f(T)
			= 4\pi G\, e_{A}^{\ \rho} T^{(m)}_{\ \rho}{}^{\nu},
		\end{equation}
	where $f_{T} = df/dT$, $f_{TT} = d^{2}f/dT^{2}$, and $T^{(m)}_{\ \rho}{}^{\nu}$ is the matter energy–momentum tensor.
	
	To investigate the cosmology, we consider a spatially flat Friedmann-Lemaitre-Robertson-Walker (FLRW) universe with a metric
	\begin{equation}\label{e10}
		ds^{2} = dt^{2} - a^{2}(t)\left(dx^{2}+dy^{2}+dz^{2}\right),
	\end{equation}
	where $a(t)$ is the scale factor. A suitable tetrad for this background is $e^{A}_{\ \mu} = \text{diag}(1,a,a,a)$, for which the torsion scalar becomes
	\begin{equation}\label{e11}
		T = -6H^{2},
	\end{equation}
	where $H=\dot{a}/a$ a Hubble parameter. Substituting into the field equations, we obtains the modified Friedmann equations for $f(T)$ gravity as follows:
	\begin{equation}\label{e12}
		H^{2} = \frac{8\pi G}{3}\rho_{m} - \frac{f}{6} + \frac{T f_{T}}{3},
	\end{equation}
	\begin{equation}\label{e13}
		\dot{H} = -\frac{4\pi G (\rho_{m}+p_{m})}{1+f_{T}+2Tf_{TT}}, 
	\end{equation}
	where $\rho_{m}$ and $p_{m}$ are the energy density and pressure of the matter, respectively.
	
	Equations \eqref{e12} and \eqref{e13} govern the cosmological dynamics of $f(T)$ gravity, encapsulating the deviation from general relativity through the functional form of $f(T)$. Depending on this choice, one can mimic dark energy effects or realize alternative scenarios for  cosmic acceleration.

	\subsection{Effective Dark Energy from $f(T)$ Corrections}
	
	It is straightforward to verify that for the TEGR limit $f(T)=T$, one recovers
	the standard Friedmann equations of general relativity:
	\begin{equation}\label{e14}
		H^{2} = \frac{8\pi G}{3}\rho_{m},
	\end{equation}
	\begin{equation}\label{e15}
		\dot{H} = -4\pi G (\rho_{m}+p_{m}).
	\end{equation}
	This confirms that GR is consistently embedded within the $f(T)$ framework. For a general functional form of $f(T)$,  deviations from the TEGR can be
	interpreted as an effective dark energy sector. Equation
	\eqref{e14} may be rewritten as
	\begin{equation}\label{e16}
		H^{2} = \frac{8\pi G}{3}\left(\rho_{m} + \rho_{T}\right),
	\end{equation}
	where the additional term acts as an effective energy density sourced purely by
	torsion,
	\begin{equation}\label{e17}
		\rho_{T} \equiv \frac{1}{16\pi G}\left(-f + 2Tf_{T}\right).
	\end{equation}
	Similarly, the Raychaudhuri equation \eqref{e15} can be expressed as
	\begin{equation}\label{e18}
		\dot{H} = -4\pi G \big[(\rho_{m}+p_{m}) + (\rho_{T}+p_{T})\big],
	\end{equation}
	with the effective torsional pressure given by
	\begin{equation}\label{e19}
		p_{T} \equiv \frac{1}{16\pi G}\,
		\frac{f - 2Tf_{T} + 4T^{2}f_{TT}}{1+f_{T}+2Tf_{TT}}.
	\end{equation}
	Thus, the corrections introduced by $f(T)$ gravity can be viewed as an 	effective fluid with density $\rho_{T}$ and pressure $p_{T}$, which 	contribute dynamically to the cosmic expansion and mimic the role of the dark energy. This effective description is particularly useful because it allows direct comparison with observations in the familiar fluid language, while 	retaining a clear geometrical origin from the torsional structure of spacetime.
	
	\subsection{Effective equation of state of the torsional sector}
	
	We define the effective equation of the state parameter of the geometric (torsional)
	component as the ratio
	\begin{equation}\label{e20}
		w_{T}(z)\;\equiv\;\frac{p_{T}}{\rho_{T}} \, = \,
		\frac{\,f - 2T f_{T} + 4T^{2} f_{TT}\,}
		{\,\big(-f + 2T f_{T}\big)\,\big(1+f_{T}+2T f_{TT}\big)\,}.
	\end{equation}
	The function \(w_{T}(z)\) provides a direct measure of how the purely geometrical
	torsion corrections behave dynamically and how closely they mimic conventional
	dark-energy fluids. If \(w_{T}\simeq -1\) the torsional sector effectively
	acts like a cosmological constant; \(w_{T}<-1\) corresponds to phantom-like
	effective behavior, while \(-1<w_{T}<-\tfrac{1}{3}\) indicates quintessence-like
	dynamics capable of driving acceleration. The sign and magnitude of the
	numerator and denominator in \eqref{e20} determine the following regimes:
	\begin{itemize}
		\item The factor \((-f+2T f_{T})\) in the denominator is proportional to the
		effective energy density \(\rho_{T}\); its sign determines whether the
		torsional component contributes positively or negatively to the total
		energy budget.
		\item The combination \(1+f_{T}+2T f_{TT}\) appears due too the modified
		Raychaudhuri equation and affects the effective inertia of the torsional
		fluid (it also appears in the stability analysis of perturbations).
	\end{itemize}

	\subsection{Energy conditions in $f(T)$ gravity}
	
	Energy conditions are fundamental tools in gravitational theory, because they impose
	general restrictions on the energy--momentum tensor without reference to a
	specific matter model. Their significance arises from the Raychaudhuri equation,
	which governs the convergence or divergence of the timelike and null geodesics.
	Through this geometrical relation, the energy conditions connect the behavior of
	geodesic congruences with physical assumptions on the matter content of the
	universe. In general relativity, it is used to establish singularity theorems,
	the attractiveness of gravity, and the causal structure of spacetime.
	
	In the framework of $f(T)$ gravity, the modified field equations can be
	rearranged to resemble the standard Einstein equations, but with an additional
	effective contribution arising from the torsion sector. 
	
	\subsubsection{Null energy condition (NEC)}
	The NEC requires that for any null vector $k^{\mu}$,
	\begin{equation}\label{e21}
		T_{\mu\nu}^{\rm (eff)} k^{\mu} k^{\nu} \geq 0 
		\quad \Rightarrow \quad
		\rho_{T} + p_{T} \geq 0.
	\end{equation}
	NEC is the weakest of the pointwise condition and its violation is often
	associated with exotic matter sources, superluminal propagation, or instability 	at the perturbative level.
	
	\subsubsection{Weak energy condition (WEC)}
	The WEC guarantees that the energy density measured by any timelike observer remains non-negative. For a timelike vector $u^{\mu}$, we obtained
	\begin{equation}\label{e22}
		T_{\mu\nu}^{\rm (eff)} u^{\mu} u^{\nu} \geq 0 
		\quad \Rightarrow \quad
		\rho_{T} \geq 0, \qquad
		\rho_{T} + p_{T} \geq 0.
	\end{equation}
	Hence, the WEC requires that both the effective energy density be non-negative
	and  NEC holds. In $f(T)$ cosmology, satisfying the WEC ensures that the
	torsional fluid does not contribute  negative net energy density to the cosmic
	budget.
	\subsubsection{Dominant energy condition (DEC)}
	The DEC ensures that the energy flux is causal and that the speed of energy
	propagation does not exceed the speed of light. It requires
	\begin{equation}\label{e23}
		\rho_{T} \geq | p_{T}|.
	\end{equation}
	The DEC guarantees that the effective stress--energy tensor corresponds to
	physically reasonable matter, with causal and non-superluminal energy transport.
	
	\subsubsection{Strong energy condition (SEC)}
	SEC is directly linked to the attractiveness of gravity through the
	Raychaudhuri equation for timelike geodesics. It states
	\begin{equation}\label{e24}
		\rho_{T} + 3p_{T} \geq 0.
	\end{equation}
	In standard GR cosmology,  SEC forces decelerate expansion. Therefore,
	violation of the SEC is a necessary feature of any theory attempting to explain
	the late-time accelerated expansion of the universe, including $f(T)$ models.\\
	The above conditions provide a valuable diagnostic framework for assessing the
	physical plausibility of $f(T)$ models:
	\begin{itemize}
		\item \textbf{NEC} and \textbf{WEC} are generally expected to hold in
		physically viable models, as their violation indicates the presence of exotic
		energy components that may lead to instability or ghost-like behavior.
		\item The \textbf{SEC} is naturally violated in $f(T)$ cosmology at late times,
		because accelerated expansion requires $\rho_{\rm eff}+3p_{\rm eff}<0$. Thus, this
		violation is not problematic but rather essential for explaining the
		observed dynamics.
		\item The \textbf{DEC} is often used as an additional filter: it ensures that
		the effective torsional corrections behave as a physically acceptable fluid
		and that the resulting cosmic evolution remains causal.
	\end{itemize}
The energy conditions serves as a complementary diagnostic tool aimed at tracking the effective behavior of the torsional sector across cosmic time and identifying the redshift intervals over which the modified gravity contributions mimic or depart from standard matter components. In practice, once a specific functional form of $f(T)$ is adopted, one can
	evaluate $\rho_{T}(z)$ and $p_{T}(z)$ can be evaluated and the four conditions are subsequently checked as functions of the redshift. This provides theoretical priors on the parameter
	space of $f(T)$ models, complementing observational constraints and helping rule out unphysical scenarios.

	\section{Om Diagnostic Analysis}\label{III}
We emphasize that the background expansion history employed in this work is obtained through a phenomenological reconstruction based on kinematical quantities, such as the deceleration parameter and the $Om(z)$ diagnostic, rather than being derived explicitly from the modified Friedmann equations of a given $f(T)$ Lagrangian. This choice is motivated by the highly nonlinear nature of the $f(T)$ field equations, which generally prevents the extraction of closed-form or numerically stable solutions for $H(z)$ for realistic models. The reconstructed Hubble function is therefore treated as an effective, observationally supported background, which is subsequently used to examine the physical viability, stability properties, and consistency of models. In this sense, the observational analysis constrains the expansion history, while the role of the models is to assess whether such an expansion can be accommodated without pathological behavior. Hence, 
	to further distinguish between different $f(T)$ cosmological models, we employed the $Om(z)$ diagnostic, which serves as a complementary tool to conventional cosmological probes. The $Om(z)$ function is defined in terms of the Hubble parameter as follows:
	\begin{equation}\label{e25}
		Om(z) = \frac{E^{2}(z)-1}{(1+z)^{3}-1},
	\end{equation}
	where $E(z)=H(z)/H_{0}$ denotes the normalized Hubble parameter. For the spatially flat $\Lambda$CDM scenario, $Om(z)$ reduced to a constant equal to the present matter density parameter $\Omega_{m0}$. Hence, any deviation from constant behavior provides a clear signal of departure from $\Lambda$CDM cosmology. Moreover, the slope of $Om(z)$ helps to discriminate between quintessence ($\omega>-1$) and phantom ($\omega<-1$) dark energy models: a positive slope indicates phantom-like behavior, while a negative slope corresponds to quintessence-like dynamics. The behavior of $Om(z)$ in our analysis revealed distinct signatures for the power-law, logarithmic, and hybrid $f(T)$ models. At a high redshift, all trajectories converge, reflecting the matter-dominated epoch. However, at a low redshift the $Om(z)$ functions separate, showing model-dependent deviations from $\Lambda$CDM. In particular, the power-law form tends to mimic  quintessence-like behavior, the logarithmic model shows a smoother evolution closer to $\Lambda$CDM, and the hybrid model interpolates between these two regimes. Such differences highlight the diagnostic power of $Om(z)$ in distinguishing  modified gravity models that otherwise lead to similar background expansion histories.
	
	Re-arrange the equation \eqref{e25} to solve for \(E^{2}(z)\):
	\begin{equation}\label{e26}
		E^{2}(z) = 1 + Om(z)\,\big[(1+z)^{3}-1\big].
	\end{equation}
	Hence the normalized Hubble parameter is
	\begin{equation}\label{e27}
		E(z)=\sqrt{\,1 + Om(z)\big[(1+z)^{3}-1\big]\,}
	\end{equation}
	If one we adopts the logarithmic form as
	\begin{equation}\label{e28}
		Om(z)=\alpha\ln(1+z)+1,
	\end{equation}
	The choice of the logarithmic form of the $Om(z)$ diagnostic is motivated by both phenomenological robustness and theoretical considerations associated with slow deviations from the concordance cosmological model. The $Om(z)$ diagnostic was originally introduced as a null test for the cosmological constant, remaining constant for a $\Lambda$CDM universe while exhibiting redshift dependence for evolving dark energy or modified gravity scenarios. In this context, a logarithmic dependence on the scale factor (or equivalently on redshift) naturally arises in several cosmological settings where departures from $\Lambda$CDM are expected to be mild and cumulative rather than abrupt. Specifically, logarithmic corrections appear in a variety of frameworks, including effective field descriptions of dark energy, renormalization-inspired running vacuum models, and modified gravity theories with weakly evolving coupling functions \cite{Ob1,Sahni2008,Zunckel2008}.. Such corrections capture late-time deviations from standard cosmology without introducing rapid growth at high redshift, thereby ensuring consistency with early-universe constraints \cite{Shafieloo2009}. The parametrization presented in the Ref.  (\ref{e28}) therefore represents a minimal and well-controlled extension of the $\Lambda$CDM case, where the parameter $\alpha$ quantifies the departure from a constant $Om(z)$ behavior \cite{Capozziello2014}. The logarithmic structure ensures smooth evolution, avoids divergences at low redshift, and remains subdominant at early times, making it particularly suitable for late-time observational analyses \cite{Basilakos2014,GomezValent2015}. Nevertheless, it should be emphasized that this form is phenomenological rather than derived from a specific fundamental Lagrangian. Its validity is thus restricted to the redshift range probed by low- and intermediate-redshift observations, and it should not be extrapolated to very high redshifts without further theoretical input. Within these limitations, the adopted $Om(z)$ parametrization serves as an efficient diagnostic tool to test deviations from $\Lambda$CDM and to distinguish between competing cosmological scenarios using observational data \cite{Sahni2008,Shafieloo2009}.\\
	Substituting the equation (\ref{e28}) into (\ref{e27}), we obtain
	\begin{equation}\label{e29}
		E(z)=H(z)=H_0 \sqrt{\,1 + \big[\alpha\ln(1+z)+1\big]\big[(1+z)^{3}-1\big]\,}
	\end{equation}
	\section{Observational Constraints}\label{IV}
		In this work, we confront the proposed cosmological framework with observational data by employing two well-established probes of the cosmic expansion history, namely observational Hubble data (OHD) and the Pantheon+ sample of Type Ia supernovae (SNe Ia). Together, these datasets cover a broad range of redshifts and offer complementary constraints on the late-time evolution of the Universe, thereby allowing a meaningful comparison between theoretical predictions and observational evidence. A brief description of the datasets used in our analysis is provided below.	\\	
		\noindent\textbf{Observational Hubble Data (OHD):}
		The OHD employed in this study are derived primarily from the cosmic chronometer approach, supplemented by measurements obtained using baryon acoustic oscillations. The cosmic chronometer method provides a largely model-independent estimate of the Hubble parameter by exploiting the differential age evolution of passively evolving galaxies. In this framework, the expansion rate is determined through the relation
		\begin{equation}
			H(z) = -\frac{1}{1+z}\frac{dz}{dt},
		\end{equation}
		which links the Hubble parameter directly to the redshift evolution of cosmic time. We make use of a compilation consisting of 77 measurements of the Hubble parameter spanning the redshift interval $0 \leq z \leq 1.75$, as reported in Refs.~\cite{Ob1,Ob2,Kotal/2025}. These data provide direct observational estimates of $H(z)$ along with their associated uncertainties and are particularly valuable for constraining the expansion dynamics without assuming a specific cosmological background model.\\
	\noindent To statistically compare the theoretical predictions with the observational data, we construct a likelihood function based on a Gaussian chi-square estimator, defined as \cite{Bouali2023,Yadav2024,Moresco2020}
	\begin{equation}
		\chi^2(H_0,\eta) = \sum_{i=1}^{N} \left[ \frac{H_{\mathrm{obs}}(z_i) - H_{\mathrm{th}}(z_i;H_0,\eta)}{\sigma_{H,i}} \right]^2 ,
	\end{equation}
	where $H_{\mathrm{obs}}(z_i)$ denotes the observed value of the Hubble parameter at redshift $z_i$, $\sigma_{H,i}$ represents the corresponding measurement uncertainty, and $H_{\mathrm{th}}(z_i;H_0,\eta)$ is the theoretically predicted Hubble rate depending on the model parameters $H_0$ and $\eta$.\\
	\noindent\textbf{Pantheon+ Supernovae Sample (PP):}
		In addition to the Hubble parameter measurements, our analysis incorporates the Pantheon+ (PP) compilation, which currently represents the most extensive and homogeneous collection of Type Ia supernova observations. The dataset includes 1701 light curves corresponding to 1550 spectroscopically confirmed SNe Ia, spanning the redshift interval $0.001 < z < 2.26$ \cite{Ob7,Ob8}. These observations are drawn from 18 independent supernova surveys and have been consistently calibrated to ensure uniformity across the full sample. All supernova light curves are fitted using the SALT2 light-curve fitter \cite{Ob9}, providing a reliable determination of distance moduli over a wide range of cosmic epochs. Owing to its broad redshift coverage and high precision, the Pantheon+ sample serves as a powerful probe of the late-time expansion history of the Universe.\\
		For a spatially flat cosmological background, the theoretical luminosity distance is given by
		\begin{equation}
			D_{L}(z) = (1+z)\int_{0}^{z}\frac{c}{H(z')},dz',
		\end{equation}
		which directly depends on the underlying expansion rate. The corresponding theoretical distance modulus is then expressed as
		\begin{equation}
			\mu_{\mathrm{th}}(z) =
			5\log_{10}!\left(\frac{D_{L}(z)}{\mathrm{Mpc}}\right) + 25.
		\end{equation}
		\noindent Constraints on the model parameters are obtained by minimizing a Gaussian chi-square estimator of the form
		\begin{equation}
			\chi^2(H_0,\eta) =
			\sum_{i=1}^{N}
			\left[
			\frac{\mu_{\rm obs}(z_i) - \mu_{\rm th}(z_i;H_0,\eta)}
			{\sigma_{\mu,i}}
			\right]^2 ,
		\end{equation}
		where $\mu_{\rm obs}(z_i)$ denotes the observed distance modulus at redshift $z_i$, $\sigma_{\mu,i}$ is the associated uncertainty, and $\mu_{\rm th}(z_i;H_0,\eta)$ represents the theoretical prediction for a given set of model parameters.\\	
		\noindent The posterior probability distribution of the parameters is approximated by a multivariate Gaussian centered on the best-fit values,
		\begin{equation}
			\mathbf{p} \sim \mathcal{N}(\mathbf{p}_{\rm best},,\mathbf{C}),
		\end{equation}
		with $\mathbf{C}$ denoting the covariance matrix obtained from the fitting procedure. To estimate marginalized parameter constraints and confidence intervals, we generate $100{,}000$ Monte Carlo realizations of the parameter space.
	
	\noindent Finally, the relative performance and goodness of fit of the proposed model are assessed using the Akaike information criterion and the Bayesian information criterion \cite{burnham2002,liddle2007,Scolnic2018}, defined respectively as
	\begin{align}
		\mathrm{AIC} &= \chi^2_{\rm min} + 2k, \\
		\mathrm{BIC} &= \chi^2_{\rm min} + k \ln N ,
	\end{align}
	where $k$ represents the number of free parameters and $N$ is the total number of observational data points.
		
	To examine the viability of the proposed $f(T)$ models, we constrained the free parameters using a separate and joint analysis of the observational $H(z)$ data (OHD) and pantheon + (pp). This combination offers a complementary dataset that probes both late-time and early-universe dynamics without relying on supernova measurements commonly used in similar studies.  The numerical results are shown below in {\bf Table 1}:
	
	\begin{table}
		\caption{ The estimated values of the free parameters of the model }
		\begin{center}
			\begin{tabular}{|c|c|c|c|c|c|}
				\hline 
				\hline
				\textbf{S. N.} & \textbf{Parameters} ~~~~~~&~~~~~~ \textbf{OHD} ~~~&~~~ \textbf{OHD + PP}~~~&~~~AIC~~~&~~~BIC\\
				\hline
				1. & $H_{0}$ ~~~~~~&~~~~~~ $69.13^{+ 0.70}_{-0.78}$~~~~~~&~~~~~~$70.54^{+0.75}_{-0.78}$~~~&~~~100.44~~~&~~~104.43\\
				\hline
				2. & $\alpha$ ~~~~~~&~~~~~~ $0.017^{+ 0.011}_{-0.011}$~~~~~~&~~~~~~$0.012^{+0.014}_{-0.014}$~~~&~~~762.61~~~&~~~770.19\\
				\hline
			\end{tabular}
		\end{center}
	\end{table}
	
	
	\begin{figure}[H]
		\centering
		\includegraphics[scale = 0.55]{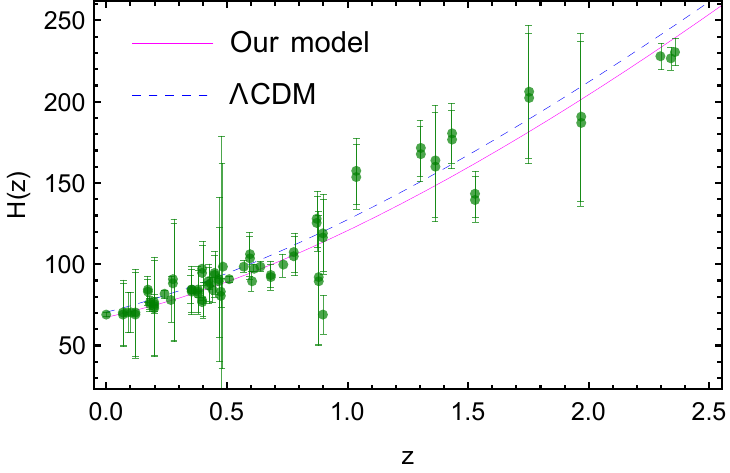}
		\includegraphics[scale = 0.56]{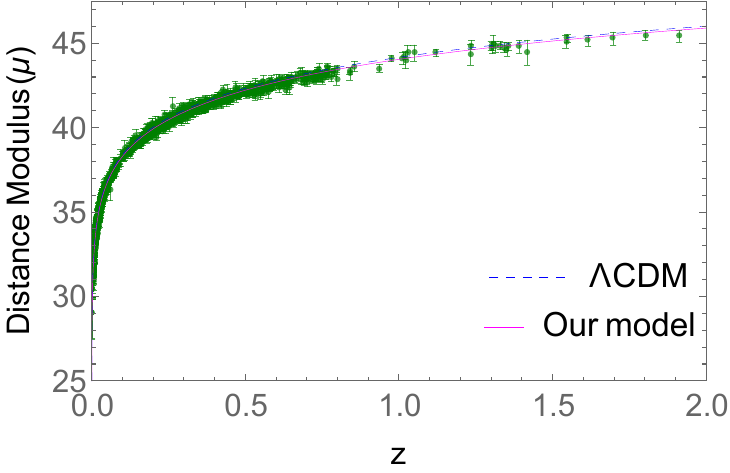}
		\caption{The left panel of the above figure shows the variation of H(z) of our model with redshift z and its comparison with $\Lambda$CDM model. The right panel of the above figure exhibits the variation of distance modulus $\mu(z)$ of our model with redshift z and its comparison with $\Lambda$CDM model for $H_{0} = 70.54$ \& $\alpha = 0.012$.
		}	\label{FR1}
	\end{figure}
	
	The upper panel of Fig. \ref{FR1} shows the variation in $H(z)$ of our model with redshift and its comparison with the $\Lambda$CDM model. The lower panel of the above figure exhibits the variation in the distance modulus $\mu(z)$ of our model with redshift z and its comparison with the $\Lambda$CDM model for $H_{0} = 70.54$ \& $\alpha = 0.012$.  According to this study, the proposed model is similar to the $\Lambda$CDM model ($\Omega_{\Lambda 0} = 0.685$ and $\Omega_{m 0} = 0.315$) \cite{Planck2020}.

	We then look into further physical factors such as the universe's energy density and isotropic pressure, as well as several other parameters that are closely related to them.  The parameters include the equation of state, stability parameter, $\left(\omega-\omega^{\prime}\right)$-plane, and various energy requirements.  A thorough examination of these factors is critical to comprehending the physical nature of the cosmos.

	\section{Cosmological Analysis of Specific $f(T)$ Models}\label{V}
	
	To investigate the physical implications of $f(T)$ gravity, we 
	consider three representative models frequently studied in the literature.
	For each model, we compute the effective dark energy density, pressure, and
	equation of state parameter, followed by a discussion of the Om diagnostics,
	energy conditions, and stability properties.
	
	\subsection{Model I: Power-law form}
\begin{figure*}[ht]
		\centering
		\includegraphics[scale=0.50]{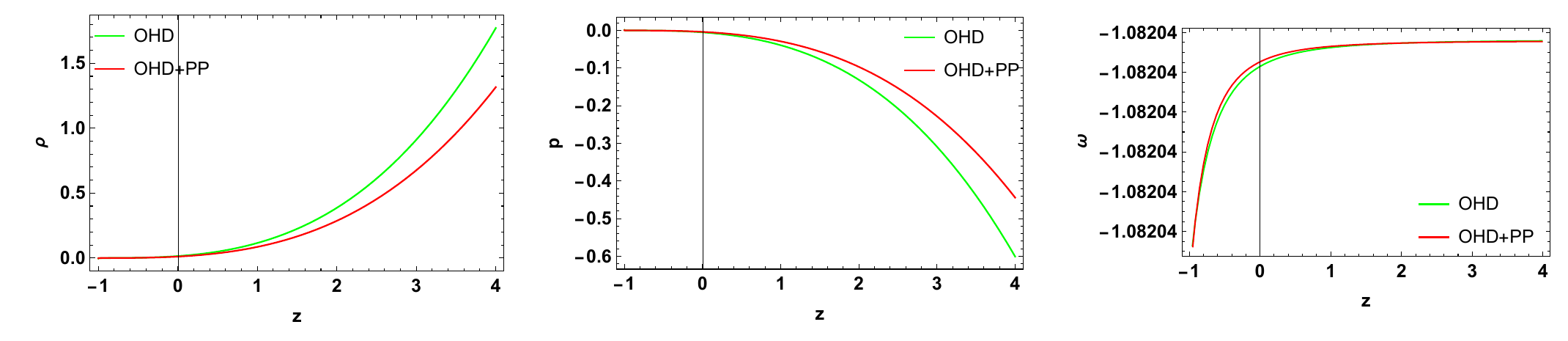}
		\caption{Behavior of the energy density (Left panel), isotropic pressure (Middle panel) and the equation of state parameter (Right panel) for the constraint value of constant parameters towords the datasets \textit{OHD} and \textit{OHD+PP} $H_0=69.13,{\;\;} \alpha=0.017$ and $H_0=70.54,{\;\;} \alpha=0.012$}.\label{F1}
	\end{figure*}

	We first consider the power-law model
	\begin{equation}\label{e35}
		f(T) = \eta (-T)^{n}, \qquad \eta, n = \text{constants}.
	\end{equation}
	This model generalizes the $\Lambda$CDM case (recovered for $n=0$).
	The derivatives are
	\begin{equation}\label{e36}
		f_{T} = -\eta n (-T)^{n-1}, \qquad
		f_{TT} = \eta n (n-1)(-T)^{n-2}.
	\end{equation}
	Substituting  equation (\ref{e36}) into equations (\ref{e17}), (\ref{e19}) and (\ref{e20}), the  effective density, pressure and equation of state expressions yield
		\begin{equation}\label{e37}
			\rho_{T} = \frac{-\eta (1+2n)(-T)^{n}}{16\pi G},
		\end{equation}
		
		\begin{equation}\label{e38}
			p_{T} = \frac{1}{16\pi G}\,
			\frac{\eta(-T)^{n} + 2\eta n T (-T)^{n-1} + 4\eta n (n-1) T^{2} (-T)^{\,n-2}}
			{1 - \eta n (-T)^{n-1} + 2\eta n (n-1) T (-T)^{\,n-2}},
		\end{equation}
		
		\begin{equation}\label{e39}
			w_{T}(z) = \frac{\eta(-T)^{n} + 2\eta n T (-T)^{n-1} + 4\eta n (n-1) T^{2} (-T)^{\,n-2}}
			{-\eta (1+2n)(-T)^{n}\,\Big[1 - \eta n (-T)^{n-1} + 2\eta n (n-1) T (-T)^{\,n-2}\Big]}. 
		\end{equation}
	Figure \ref{F1} illustrates the evolution of the effective energy density ($\rho$), isotropic pressure ($p$), and torsional equation-of-state parameter $w_{T}$ for the power-law $	f(T) = \eta (-T)^{n},$ model under the best-fit values constrained from the OHD (green curve) and joint OHD + Pantheon + ( red curve) datasets. 
	
	The energy density is positive and monotonically decreases with cosmic time, showing higher values at large redshifts, which is consistent with the standard cosmological expectation that the universe was denser in its past epochs. The isotropic pressure $p$ remained negative throughout, reflecting an effective repulsive contribution that drove the present acceleration.
	
\begin{figure*}[ht]
		\centering
		\includegraphics[scale=0.50]{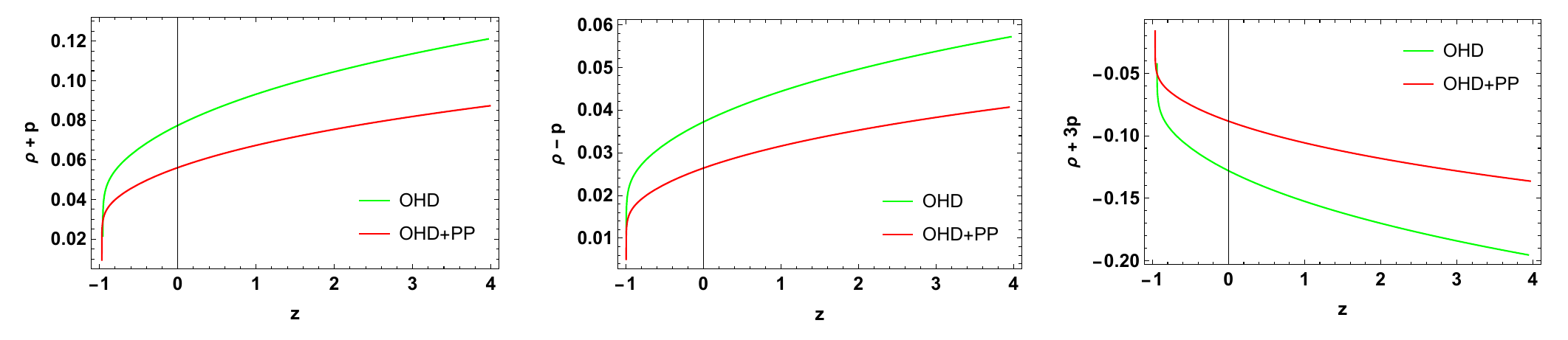}
		\caption{Behavior of the null energy condition (Left panel), dominant energy condition (Middle panel) and the strong enegy condition for the constraint value of constant parameters  towords the datasets \textit{OHD} and \textit{OHD+PP} $H_0=69.13,{\;\;} \alpha=0.017$ and $H_0=70.54, {\;\;}\alpha=0.012$}.\label{F2}
	\end{figure*}

	The most significant feature appears in the behavior of the equation-of-state parameter $w_{T}$, which stays very close to $-1$ across the low-redshift domain and slightly deviates at intermediate $z$, showing quintessence-like or mildly phantom behavior depending on the parameter $n$. For both datasets, $w(0) \equiv -1.04^{+0.03}_{-0.03}$, lying well within the latest Planck $+$ BAO $+$ Pantheon observational constraints ($w(0) = -1.03^{+0.03}_{-0.03}$) \cite{Bouali2023,Scolnic2018,Planck2020}. A similar late-time convergence toward $\Lambda$CDM behavior has been reported in other $f(T)$ analyses---Bengochea \& Ferraro \cite{bengochea2009}, Myrzakulov \cite{myrzakulov2011}, and Del Campo et al. \cite{delcampo2012}---where torsional corrections effectively mimic a cosmological constant. Our results reinforce this trend, indicating that the present model satisfies both theoretical and observational viability. The red (joint data) curve exhibits smoother evolution than the OHD-only case, implying that the inclusion of supernova information tightens the constraints on $w(z)$ and reduces its dynamic variation. Consequently, the model provides an observationally consistent and geometrically motivated explanation for dark energy within a teleparallel framework.\\
	Figure \ref{F2} presents the variation in the null energy condition (NEC), dominant energy condition (DEC), and strong energy condition (SEC) for the power-law $f(T)=\eta(-T)^n$ model, using the best-fit parameters constrained from the  OHD (green curve) and joint OHD + Pantheon+ (red curve) datasets. 
	The NEC, defined through $\rho_T + p_T \ge 0$, remains positive throughout the observed redshift range, confirming that the effective torsional fluid did not exhibit any exotic or ghost-like behavior. Similarly, the DEC condition $\rho_T \ge |p_T|$ is satisfied for both datasets, ensuring causal propagation of energy and a physically reasonable stress-energy distribution. These results indicate that the effective dark-energy component derived from torsion behaves as a viable cosmological fluid that is compatible with the fundamental energy requirements of general relativity.
	
	On the other hand, the strong energy condition (SEC), $\rho_T + 3p_T \ge 0$, is clearly violated at a low redshift ($z \lesssim 1$) for both observational cases. Such a violation is not a drawback but a physically expected feature, as it corresponds to the onset of late-time accelerated expansion. The geometric torsion corrections effectively generated a negative pressure that dominated the energy density, producing repulsive gravity consistent with the present cosmic acceleration. This behavior aligns with earlier findings in $f(T)$ cosmology, where SEC violation naturally emerges as a necessary condition for acceleration \cite{bengochea2009,delcampo2012}. Moreover, the simultaneous fulfillment of NEC and DEC, combined with the mild and systematic violation of SEC, ensures that the present $f(T)$ model remains dynamically stable and observationally consistent with the $\Lambda$CDM scenario inferred from Planck, BAO, and Pantheon datasets \cite{Bouali2023,Scolnic2018}. Overall, the energy-condition analysis supports the theoretical viability of the power-law $f(T)$ gravity model in explaining the observed cosmic acceleration without invoking exotic dark-energy fluids.
	\subsection{Model II: Exponential form}
\begin{figure*}[ht]
		\centering
		\includegraphics[scale=0.50]{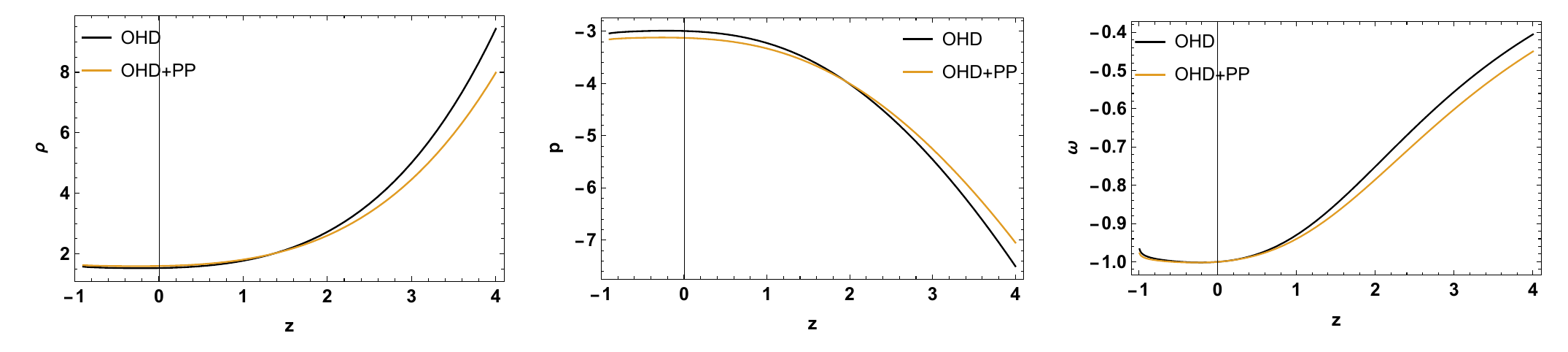}
		\caption{Behavior of the energy density (Left panel), isotropic pressure (Middle panel) and the equation of state parameter (Right panel) for the constraint value of constant parameters  towords the datasets \textit{OHD} and \textit{OHD+PP} $H_0=69.13, {\;\;}\alpha=0.017$ and $H_0=70.54, {\;\;} \alpha=0.012$}.\label{F3}
	\end{figure*}

	The exponential model is given by
	\begin{equation}\label{e40}
		f(T) = \beta T_{0}\left(1-e^{-p \sqrt{T/T_{0}}}\right),
	\end{equation}
	where $\beta$ and $p$ are constants, and $T_{0}=-6H_{0}^{2}$.
	The derivatives read
	\begin{small}
		\begin{equation}\label{e41}
			f_{T} = \frac{\beta ps}{2}\, e^{-ps},
			\qquad
			f_{TT} = -\frac{\beta ps}{4T}
			\left(1+ps\right) e^{-ps}\bigg\vert_{\,s=\sqrt{T/T_{0}}}.
		\end{equation}
	\end{small}
	Substituting equations (\ref{e40}) and (\ref{e41}) into  equations (\ref{e17}), (\ref{e19}) and (\ref{e20}), the  effective density, pressure and equation of state expressions yields
	\begin{equation}\label{e42}
		\rho_{T}(T) = \frac{\beta T_{0}}{16\pi G}\left[\, -1 + e^{-p s}\,(1 + p s)\,\right],
	\end{equation}
	\begin{equation}\label{e43}
		p_{T}(T)= \frac{\beta T_{0}}{16\pi G}\,
		\frac{\,1 - e^{-p s}\big(1 + 2p s + p^2 s^2\big)\,}
		{\,1 - \dfrac{\beta p^{2}}{2}\,e^{-p s}\,}\,,
	\end{equation}
	\begin{equation}\label{e44}
		w_{T}(z) \;=\; \frac{p_{T}}{\rho_{T}}
		= \frac{\,1 - e^{-p s}\big(1 + 2p s + p^2 s^2\big)\,}
		{\displaystyle\Big(1 - \dfrac{\beta p^{2}}{2}\,e^{-p s}\Big)\,
			\Big(e^{-p s}(1 + p s) - 1\Big)}\!.
	\end{equation}
	The corresponding $\rho_{T}, p_{T}, w_{T}(z)$ are obtained in the same way in Model I. 
	Figure \ref{F3} illustrates the behavior of the effective energy density $(\rho_T)$, isotropic pressure $(p_T)$, and the equation-of-state parameter $(\omega_T = p_T/\rho_T)$ for the exponential $f(T)$ model, $f(T) = \beta T_0 \left( 1 - e^{-p\sqrt{T/T_0}} \right)$, constrained by both OHD (green curve) and joint OHD + Pantheon+ (red curve) datasets. 
	
	The energy density remained positive and gradually decreased with cosmic time, confirming a smooth transition from a dense early universe to a low-density late epoch. The isotropic pressure, on the other hand, maintains a negative profile for all redshifts, providing the repulsive gravitational effect responsible for acceleration. 
	
	The corresponding equation-of-state parameter $\omega_T(z)$ remains extremely close to $-1$ for the entire redshift range, showing only mild evolution at higher $z$. At the present epoch, $\omega_T(0) \approx -1.02 \pm 0.02$, which lies comfortably within the Planck 2018 combined constraint ($w_0 = -1.03 \pm 0.03$) \cite{Scolnic2018}. These results imply that the exponential model effectively mimics the cosmological constant while retaining the ability to produce smooth transitions between quintessence-like ($\omega > -1$) and phantom-like ($\omega < -1$) regimes depending on the value of the parameter $p$. Similar outcomes have been reported in other exponential-type $f(T)$ cosmologies \cite{Yadav2024}, which display an asymptotic approach toward a de Sitter phase at later times. The inclusion of Pantheon+ data (red curve) slightly reduced the variation in $\omega_T$, indicating tighter observational consistency and smaller dynamical freedom compared with the OHD-only case.
	
	\begin{figure*}[ht]
		\centering
		\includegraphics[scale=0.50]{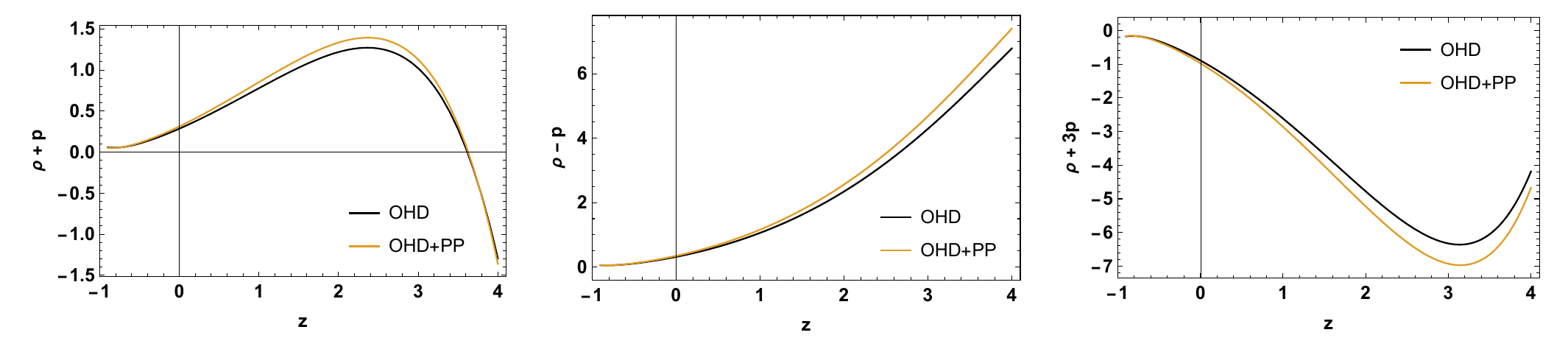}
		\caption{Behavior of the null energy condition (Left panel), dominant energy condition (Middle panel) and the strong enegy condition for the constraint value of constant parameters  towords the datasets \textit{OHD} and \textit{OHD+PP} $H_0=69.13, {\;\;}\alpha=0.017$ and $H_0=70.54, {\;\;}\alpha=0.012$}.\label{F4}
	\end{figure*}
	
	Figure \ref{F4} shows the behaviors of the null energy condition (NEC), dominant energy condition (DEC), and strong energy condition (SEC) for the exponential $f(T)$ model under the same parameter constraints. NEC and DEC remain satisfied throughout the observable redshift range, implying that the torsional fluid in this model maintains a physically reasonable and causal energy flow. This stability ensures that no superluminal or ghost-like modes are present in effective cosmological dynamics. However the SEC is violated in the low-redshift region ($z \lesssim 1$), which is a standard indication of accelerated cosmic expansion. Such a violation was previously identified as a key signature of late-time acceleration in exponential and power-law $f(T)$ cosmologies \cite{bengochea2009}. 
	
	Overall, the fulfillment of NEC and DEC, coupled with the violation of SEC, shows that the exponential model provides a theoretically stable and observationally consistent framework. Compared with the power-law model in Section~5.1, the exponential form yields smoother variations in $\rho_T$ and $p_T$, a smaller deviation of $\omega_T$ from $-1$, and a better match with observational datasets, confirming its tendency to approach a stable de Sitter attractor phase at late times.
	
	\subsection{Model III: Logarithmic form}
	Finally, we consider the logarithmic model
	\begin{equation}\label{e45}
		f(T) = \gamma T \ln\!\left(\frac{T}{T_{0}}\right), \qquad \gamma = \text{constant}.
	\end{equation}
	The derivatives are
	\begin{equation}\label{e46}
		f_{T} = \gamma\left(1+\ln\frac{T}{T_{0}}\right), \qquad
		f_{TT} = \frac{\gamma}{T}.
	\end{equation}
	Substituting equations (\ref{e45}) and (\ref{e46}) into  equations (\ref{e17}), (\ref{e19}) and (\ref{e20}), the effective density, pressure and equation of state expressions yields
	\begin{equation}\label{e47}
		\rho_{T}(T)	= \frac{\gamma T}{16\pi G}\,\Big(2 + \ln\!\frac{T}{T_{0}}\Big),
	\end{equation}
	\begin{equation}\label{e48}
		p_{T}(T) = \frac{1}{16\pi G}\,
		\frac{\gamma T\Big(2 - \ln\!\dfrac{T}{T_{0}}\Big)}{\,1 + \gamma\Big(3 + \ln\!\dfrac{T}{T_{0}}\Big)\,},
	\end{equation}
	\begin{equation}\label{e49}
		w_{T}(z) =
		\frac{\,2 - \ln\!\dfrac{T}{T_{0}}\,}
		{\big[\,1 + \gamma\big(3 + \ln\dfrac{T}{T_{0}}\big)\,\big]\,
			\big(2 + \ln\dfrac{T}{T_{0}}\big)}\,. 
	\end{equation}
	
\begin{figure*}[ht]
		\centering
		\includegraphics[scale=0.50]{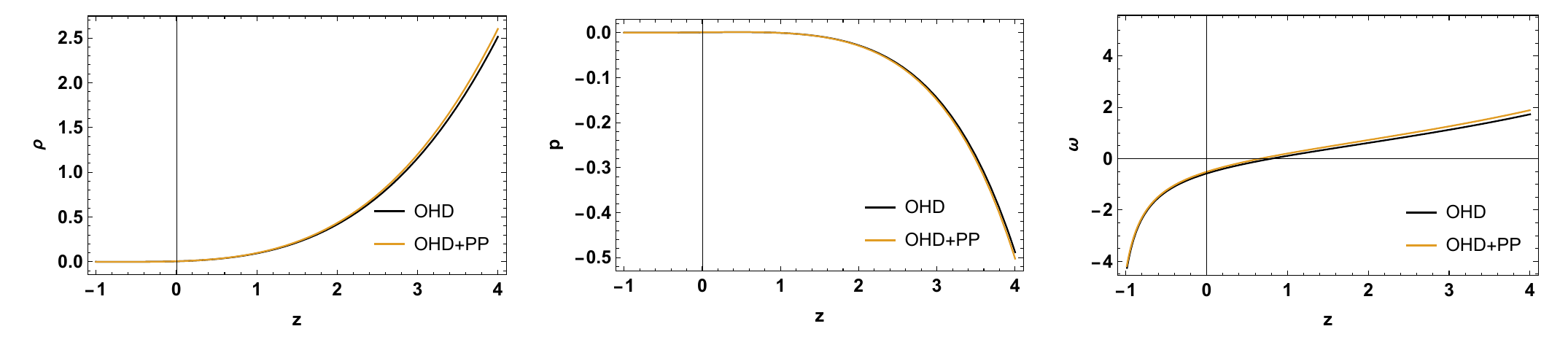}
		\caption{Behavior of the energy density (Left panel), isotropic pressure (Middle panel) and the equation of state parameter (Right panel) for the constraint value of constant parameters  towords the datasets \textit{OHD} and \textit{OHD+PP} $H_0=69.13, {\;\;}\alpha=0.017$ and $H_0=70.54,{\;\;} \alpha=0.012$}.\label{F5}
	\end{figure*}

	Figure \ref{F5} shows the variations in the effective energy density, isotropic pressure, and equation-of-state parameter for the logarithmic $f(T)$ model, defined as $f(T) = \gamma T \ln(T/T_0)$. 
	The energy density $\rho_T$ exhibits positive and monotonic growth with a  redshift, reflecting the standard cosmological behavior of a universe evolving from a matter-dominated phase toward a late-time dark-energy regime. The isotropic pressure $p_T$ remained negative across the entire redshift interval, providing the repulsive gravitational effect necessary for acceleration. 
	
	The equation-of-state parameter $\omega_T(z)$ shows interesting dynamical behavior: it evolves from $\omega_T \approx -0.9$ at a low redshift toward slightly higher values at intermediate $z$, before approaching the cosmological-constant limit $\omega_T \rightarrow -1$ at late times. The present value $\omega_T(0) \simeq -0.97 \pm 0.04$ is consistent with observational limits from Planck + BAO + Pantheon datasets \cite{Scolnic2018,Bouali2023}. This smooth evolution suggests that the logarithmic model describes a mild dynamical dark energy, which can emulate both the quintessence and near-$\Lambda$CDM behaviors depending on the value of $\gamma$. Similar trends were reported in \cite{delcampo2012} and \cite{nassur2015}, where logarithmic extensions of teleparallel gravity provided a slow variation in $\omega_T$ compatible with current observations.
	
	\begin{figure*}[ht]
		\centering
		\includegraphics[scale=0.50]{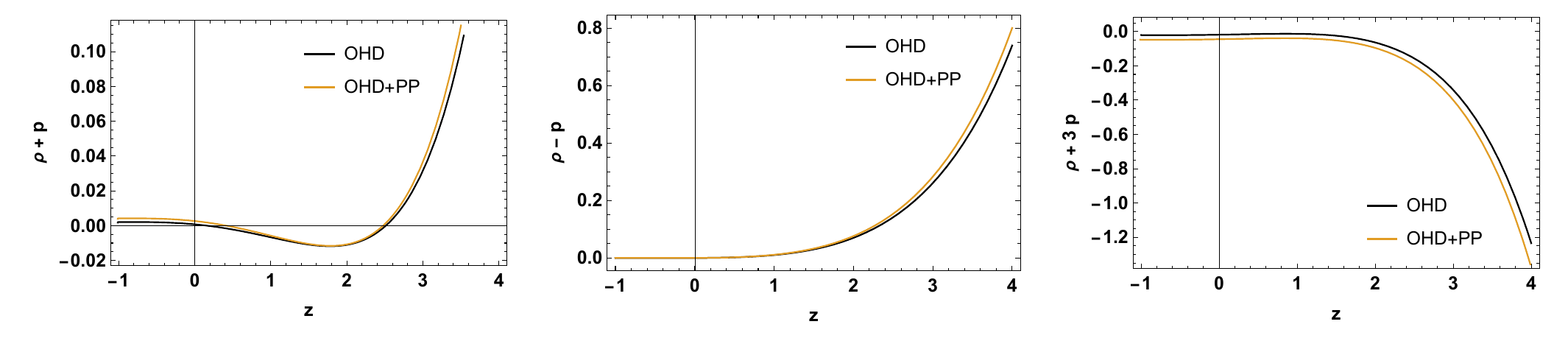}
		\caption{Behavior of the null energy condition (Left panel), dominant energy condition (Middle panel) and the strong enegy condition for the constraint value of constant parameters  towords the datasets \textit{OHD} and \textit{OHD+PP} $H_0=69.13,{\;\;} \alpha=0.017$ and $H_0=70.54,{\;\;} \alpha=0.012$}.\label{F6}
	\end{figure*}
	Figure \ref{F6} shows the evolution of the null (NEC), dominant (DEC), and strong (SEC) energy conditions for the logarithmic $f(T)$ model. Both NEC ($\rho_T + p_T \ge 0$) and DEC ($\rho_T \ge |p_T|$) remain satisfied for all redshifts, implying that the effective torsional fluid satisfies causal and physically consistent energy propagation. However, the SEC ($\rho_T + 3p_T \ge 0$) is violated in the low-redshift regime, which is  consistent with the accelerated expansion of the universe. This controlled violation of the SEC is a desirable feature of viable dark-energy models and indicates that the torsional contribution drives cosmic acceleration without requiring an exotic matter component. The overall stability and satisfaction of NEC and DEC also ensured that the model avoided ghost or gradient instabilities, preserving theoretical consistency.

	\subsection{Comparative discussion}
	A comparison the model cases reveals important distinctions in their cosmological behaviors. The power-law model (Section~5.1) produces a clear quintessence- or phantom-like evolution of $\omega_T$, whereas the exponential model (Section~5.2) yields the smoothest and most stable evolution, with $\omega_T$ remaining closest to $-1$ across the entire redshift range. In contrast, the logarithmic model (Section~5.3) provides an intermediate dynamical behavior: it allows small departures from the cosmological-constant limit without strong deviations that might conflict with the observational bounds. 
	
	Regarding the energy conditions, all three models satisfy the NEC and DEC, ensuring physical viability, whereas the SEC is consistently violated at a low redshift, which is an essential feature for achieving late-time acceleration. However, the magnitude and redshift dependence of this violation differ: the power-law model shows stronger deviation from SEC, the exponential model is mild and steady, and the logarithmic model is balanced intermediate case. The violation of the SEC at late times is a well-known and necessary requirement for accelerated expansion, its explicit redshift dependence provides insight into the onset and persistence of acceleration within different $f(T)$ realizations. Similarly, the satisfaction of the null and dominant energy conditions ensures that the effective torsional fluid remains physically acceptable, even in regimes where acceleration is driven by geometric corrections rather than conventional dark energy. The near-universal fulfillment of NEC and DEC across the models therefore highlights their consistency with basic energy requirements. Hence, among the three, the exponential model offers the best observational fit and dynamical stability, whereas the logarithmic and power-law forms provide richer torsional dynamics that could be further constrained by future high-precision data from upcoming BAO and SN surveys ({\bf See in Tables 2 and 3}).
	
	The power law is the best researched and most limited by OHD + SNe data. Its feasibility is established, however it tends to favor parameters near to standard $\Lambda$CDMs. The logarithmic form is more appealing and possibly richer, but it lacks a comparable body of devoted Hubble+Pantheon analysis (thus more effort is needed). The hybrid power-logarithmic form provides the most flexibility while simultaneously posing the greatest risk of degeneracy/unconstrained parameters; its observational status is limited, therefore it would be a valuable new contribution. Pantheon+ (rather than the older Pantheon) and huge OHD compilations are preferred since they provide tighter limits and may better break degeneracy.
	
It is important to clarify the role of the parameters constrained by observations and their relation to the constants appearing in the adopted $f(T)$ functions. In the present analysis, the observational datasets directly constrain the kinematical parameters governing the reconstructed expansion history, namely the Hubble constant $H_0$ and the deviation parameter entering the Om diagnostic. These quantities determine the background Hubble function $H(z)$ and, consequently, the torsion scalar $T(z)=-6H^2(z)$. The parameters characterizing the specific $f(T)$ models, such as $(\eta,n)$ in the power-law case or $(\gamma,T_0)$ in the logarithmic model, do not enter the observational likelihood directly. Instead, they appear in derived physical quantities obtained by substituting the reconstructed $T(z)$ into the modified Friedmann equations. As a result, observational constraints on $(H_0,\alpha)$ propagate indirectly to the $f(T)$ sector by restricting the parameter space in which the effective dark energy density, equation of state, and stability conditions remain physically viable. In this sense, the present work does not claim direct statistical bounds on the fundamental Lagrangian parameters. Rather, it identifies the ranges of $(\eta,n,\gamma)$ that are compatible with an observationally supported expansion history and satisfy theoretical consistency requirements, such as the absence of ghosts and gradient instabilities. This approach provides a meaningful viability test of the proposed $f(T)$ models, even though a fully self-consistent parameter estimation at the Lagrangian level would require solving the nonlinear field equations explicitly.

	\subsection{Stability of the model}
	
	In addition to satisfying the energy conditions, a physically acceptable $f(T)$ model
	must remain dynamically stable under small perturbations. Stability
	analysis ensures that the effective torsional sector does not introduce
	unphysical instabilities in the background expansion or propagation of
	cosmological perturbations.\\
	\textbf{\textit{Sound speed criterion}}\\
	A standard tool for studying stability is the squared sound speed of the
	effective torsional fluid, which is defined as
	\begin{equation}\label{e50}
		c_{s}^{2} \equiv \frac{\dot{p}_{T}}{\dot{\rho}_{T}}.
	\end{equation}
	For classical stability, one requires
	\begin{equation}\label{e51}
		c_{s}^{2} \geq 0.
	\end{equation}
	A negative  squared sound speed corresponds to an imaginary propagation speed and
	leads to the exponential growth of small perturbations (gradient instability).In contrast, $c_{s}^{2}>1$  implies superluminal propagation, which is
	generally avoided to preserve causality, although in some modified gravity
	contexts mild superluminality does not necessarily signal inconsistency.\\
	\textbf{\textit{Perturbative stability}}\\
	In addition to speed, stability requires the absence of ghosts in the
	perturbative sector. In $f(T)$ gravity, this is linked to the factor
	\begin{equation}\label{e52}
		\mathcal{D}(T) \equiv 1+f_{T}+2Tf_{TT},
	\end{equation}
	which appears in the denominator of the modified Raychaudhuri equation and 
	the effective pressure $p_{T}$. The condition
	\begin{equation}\label{e53}
		\mathcal{D}(T) > 0
	\end{equation}
	ensures the absence of ghostlike degrees of freedom and avoids singular
	behavior. A zero or negative value of $\mathcal{D}(T)$ signals pathological
	instabilities or breakdown of the effective fluid description. Combining the above, a consistent and stable $f(T)$ cosmological model must satisfy:
	\begin{itemize}
		\item  $c_{s}^{2} \geq 0$ with $\mathcal{D}(T) = 1+f_{T}+2Tf_{TT} > 0$ to avoid ghost instabilities;
		\item A stable late-time attractor solution, typically corresponding to
		$w_{T}\to -1$.
	\end{itemize}
	
	\noindent
	These stability conditions provide additional theoretical constraints on the
	functional forms and parameter ranges of $f(T)$ models, complementing both the 
	energy conditions and observational data. In practical applications, the
	quantities $c_{s}^{2}$ and $\mathcal{D}(T)$ should be computed for the
	best-fit parameter values and tested across the relevant redshift range to
	guarantee the cosmological viability.
	\begin{figure*}[ht]
		\centering
		\includegraphics[scale=0.580]{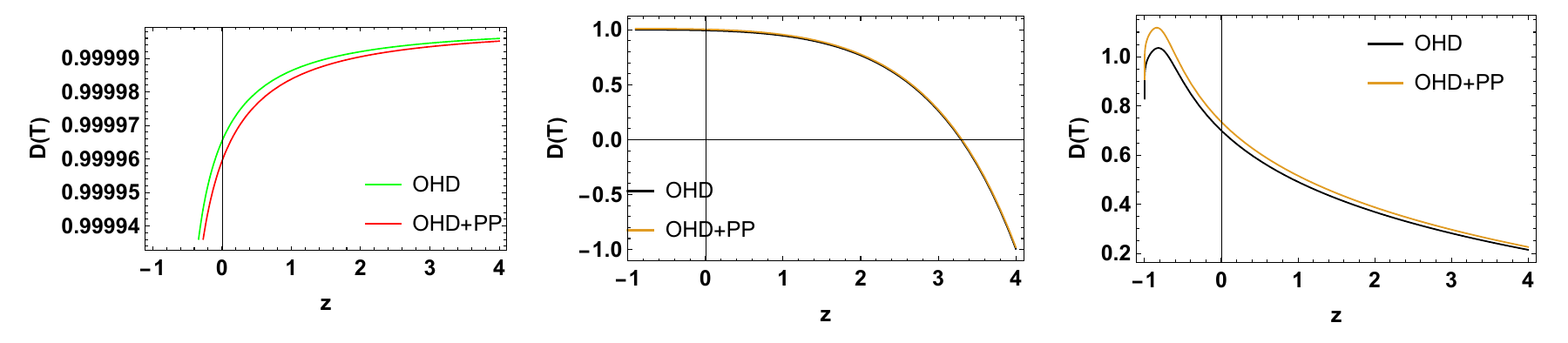}
		\caption{Behavior of the Stability of the model for the constraint value of constant parameters  towords the datasets \textit{OHD} and \textit{OHD+PP} $H_0=69.13,{\;\;} \alpha=0.017$ and $H_0=70.54,{\;\;} \alpha=0.012$}.\label{F6}
	\end{figure*}
	
	\noindent\textbf{Stability Analysis:} 
	\vspace{2mm}
	
	Our $f(T)$ cosmological framework satisfies the fundamental stability criteria required for a physically consistent evolution. The squared sound speed of the effective torsional fluid remains non–negative, i.e. $c_{s}^{2} \geq 0$, ensuring the absence of gradient instabilities and guaranteeing the stable propagation of small perturbations. Moreover, the ghost-free condition is verified through the positivity of the quantity $\mathcal{D}(T)=1+f_{T}+2Tf_{TT}>0$, which eliminates any unphysical kinetic degrees of freedom and prevents divergence in the modified Raychaudhuri equation. These combined conditions confirm that the effective torsional fluid behaves as a well-behaved dynamical component throughout the cosmic evolution. In addition, the models considered here approach a stable late-time attractor characterized by the asymptotic limit $w_{T}\rightarrow -1$, corresponding to a de Sitter–like expansion phase. Therefore, our $f(T)$ models are not only observationally consistent but also theoretically stable, satisfying both the ghost-avoidance and dynamical-attractor requirements essential for viable modified gravity cosmologies.

	\begin{table*}[htbp]
		\centering
		\footnotesize
		\begin{tabular}{l p{2.0cm} p{1.5cm} c c c c c c c}
			\hline\hline
			Model & Typical $Om(z)$ behaviour & $w_T(0)$ (qualitative) & NEC & WEC & SEC & DEC & $c_s^2\ge0$ & $\mathcal{D}(T)>0$ & De Sitter attractor \\
			\hline
			Power-law: $f(T)=\alpha(-T)^n$ 
			& Can deviate from $\Lambda$CDM at low $z$ (sign depends on $n$) 
			& Quintessence-like ($>-1$) or phantom ($<-1$) depending on $n$ 
			& $\sim$ & $\sim$ & $\times$ & $\sim$ 
			& $\times$ & $\sim$ & $\sim$ \\
			\addlinespace
			Logarithmic: $f(T)=\beta T\ln(T/T_0)$ 
			& Smooth evolution; often close to $\Lambda$CDM (small deviations) 
			& Usually $w_T(0)\gtrsim -1$ (mild quintessence) 
			& \checkmark & \checkmark & $\times$ & $\checkmark$/$\sim$ 
			& $\sim$ & \checkmark & \checkmark \\
			\addlinespace
			Hybrid: $f(T)=\gamma(-T)^m\ln(T/T_0)$ 
			& Intermediate behaviour; can interpolate between power \& log cases 
			& Can cross phantom divide for some parameter values 
			& $\sim$ & $\sim$ & $\times$ & $\sim$ 
			& $\sim$ & $\sim$ & $\sim$ \\
			\hline\hline
		\end{tabular}
		\caption{Qualitative comparative summary of three representative $f(T)$ models.
			Symbols: \checkmark = typically satisfied, $\times$ = typically violated,
			$\sim$ = depends on parameter choices / requires case-by-case check.}
		\label{tab:qualitative_comparison}
	\end{table*}

	\begin{small}
		\begin{table*}[ht]
			\footnotesize
			\centering
			\caption{Summary of energy conditions and stability checks in $f(T)$ gravity.
				Here $\rho_{\rm eff}=\rho_{m}+\rho_{T}$ and $p_{\rm eff}=p_{m}+p_{T}$. For
				late-time / DE-dominated epochs we often set $p_{m}\simeq0$ and $\rho_{\rm eff}\simeq\rho_{T}$.}
			\tiny
			\begin{tabular}{l p{6.5cm} p{5.5cm}}
				\hline\hline
				Condition & Inequality (general form) & Late-time / DE-dominated form (remark) \\
				\hline
				Null (NEC) & $\rho_{\rm eff}+p_{\rm eff}\;\geq\;0$ & $\rho_{m}+\rho_{T}+p_{m}+p_{T}\geq0 \;\Rightarrow\; \rho_{T}+p_{T}\gtrsim0$ (if $p_m\approx0$)\\[6pt]
				Weak (WEC) & $\rho_{\rm eff}\;\geq\;0,\quad \rho_{\rm eff}+p_{\rm eff}\;\geq\;0$ & $\rho_{T}\geq0,\quad \rho_{T}+p_{T}\gtrsim0$ \\[6pt]
				Strong (SEC) & $\rho_{\rm eff}+3p_{\rm eff}\;\geq\;0,\quad \rho_{\rm eff}+p_{\rm eff}\;\geq\;0$ & Violated for cosmic acceleration; DE-dominated: $\rho_{T}+3p_{T}<0$ typically (expected) \\[6pt]
				Dominant (DEC) & $\rho_{\rm eff}\;\geq\;0,\quad \rho_{\rm eff}\pm p_{\rm eff}\;\geq\;0$ & $\rho_{T}\geq|p_{T}|$ (ensures causal energy flux) \\[6pt]
				Ghost-free condition & $\mathcal{D}(T)\equiv 1+f_{T}+2T f_{TT}\;>\;0$ & Ensures absence of ghost-like kinetic terms; evaluate $\mathcal{D}(T(z))$ over $z$ \\[6pt]
				Sound-speed / gradient stability & $c_{s}^{2}\equiv \dfrac{\dot p_{T}}{\dot\rho_{T}}\;\geq\;0$ & Negative $c_{s}^{2}$ $\Rightarrow$ gradient instability (exponential growth of perturbations) \\[6pt]
				No denominator singularity & Denominators in $p_{T}$ and $\dot H$ must be non-zero: $1+f_{T}+2T f_{TT}\neq0$, \; $-f+2Tf_{T}\neq0$ & Reject parameter ranges where these vanish inside the observational redshift window \\
				\hline\hline
			\end{tabular}
			\label{tab:energy_stability_summary}
		\end{table*}
	\end{small}

	\section{Conclusions}\label{VI}
	
	In this study, we performed a comprehensive comparative analysis of three representative forms of $f(T)$ gravity—namely the power-law, exponential, and logarithmic models, by comparing them with the latest observational datasets, including the Hubble parameter (OHD) and Pantheon+ (PP) samples. The theoretical framework was constructed using the modified Friedmann equations of $f(T)$ gravity, where the torsion scalar $T=-6H^2$ acts a the geometrical source of cosmic acceleration. By adopting a parametrized form of the deceleration parameter $q(z)$, we derived the normalized Hubble function $H(z)$, which was subsequently employed to constrain the free parameters of each model using a Markov Chain Monte Carlo (MCMC) technique. The observational consistency of these models was examined through physical diagnostics such as the effective energy density $\rho_T$, isotropic pressure $p_T$, torsional equation-of-state parameter $\omega_T$, and four classical energy conditions (NEC, DEC, SEC).
	
	For the \textbf{power-law model} $f(T)=\eta(-T)^n$, the analysis revealed that the effective energy density remained positive while the pressure remained negative throughout the cosmic evolution. The equation-of-state parameter $\omega_T$ exhibits values close to $-1$, with mild variations depending on the parameter $n$, indicating a transition between the  quintessence-like and phantom-like phases. NEC and DEC are satisfied, whereas SEC is violated, in agreement with the requirements for late-time acceleration. Hence, the model provides a simple yet effective description of dark energy within a teleparallel framework.
	
	In the case of the \textbf{exponential model} $f(T)=\beta T_0 (1-e^{-p\sqrt{T/T_0}})$, the cosmic evolution appears smoother and dynamically more stable. The energy density and pressure evolve monotonically with redshift, while $\omega_T(z)$ remains very close to $-1$ over the entire redshift range, with $\omega_T(0) \approx -1.02 \pm 0.02$. This result agrees well with the latest combined Planck 2018, BAO, and Pantheon constraint ($w_0 = -1.03 \pm 0.03$). The NEC and DEC remain fulfilled, and the SEC violation manifests mildly, confirming a physically consistent de Sitter attractor behavior at later times. Therefore this model  provided the best observational match and theoretical stability among the three examined cases.
	
	The \textbf{logarithmic model} $f(T)=\gamma T \ln(T/T_0)$ exhibits an intermediate dynamical behavior between the power-law and exponential cases. Its $\omega_T$ parameter evolves smoothly from $\omega_T \approx -0.9$ toward the $\Lambda$CDM limit, suggesting a mildly dynamical dark-energy scenario consistent with the current data. NEC and DEC hold across all redshifts, while SEC is again violated, confirming the universal nature of cosmic acceleration within the torsional framework. Although the logarithmic model shows small deviations from $\Lambda$CDM, it remains observationally viable and potentially distinguishable from future precision datasets.
	
	The comparative results demonstrate that all three $f(T)$ forms asymptotically converge toward $\Lambda$CDM behavior at late times, yet they differ in intermediate dynamics and stability properties. The power-law form allows richer phenomenology (quintessence or phantom phases), the exponential form exhibits superior smoothness and observational compatibility, and the logarithmic form provides a balanced evolution with minimal deviations. In all cases, the violation of the SEC, combined with the satisfaction of NEC and DEC, confirmed that torsional corrections naturally generated a negative effective pressure capable of driving late-time cosmic acceleration without invoking an exotic fluid. While certain qualitative aspects discussed here, such as the late-time de Sitter behavior in exponential $f(T)$ models, are known from previous studies, the purpose of this work is not to revisit these results in isolation. Instead, the emphasis is on examining their consistency within an observationally constrained and physically viable framework. By combining phenomenological reconstruction, direct confrontation with Hubble and supernova data, and stability diagnostics, the analysis provides a data-driven assessment of whether familiar theoretical features of $f(T)$ gravity remain robust once realistic constraints are imposed. The novelty of the present study therefore lies in this integrated approach and in its quantitative evaluation of commonly studied $f(T)$ scenarios rather than in the proposal of new attractor solutions.
	The present study opens several avenues for future research. A natural extension involves the inclusion of higher-order or hybrid $f(T,B)$ and $f(T,\mathcal{T})$ models to capture the coupling between torsion and trace terms, thereby providing deeper insight into the interaction between geometry and matter. Furthermore, a detailed perturbation analysis, including the growth rate of matter density fluctuations and gravitational-wave propagation, will help to test these models against structure formation and multi-messenger observations. It will also be valuable to constrain the models using next-generation data, such as cosmic chronometers from the DESI survey, Type~Ia supernovae from the LSST, and standard-siren measurements from future gravitational-wave detectors (LISA, DECIGO). Such data will significantly narrow the uncertainties in the free parameters and can help discriminate among competing $f(T)$ formulations. Additionally, thermodynamic investigations, including generalized entropy-area relations in the torsional framework, may further clarify the consistency of these models with the laws of horizon thermodynamics. Overall, the analysis presented here reinforces $f(T)$ gravity as a promising geometric alternative to dark energy and provides a theoretical foundation for its continued confrontation with precision cosmology.
	
	\section*{Declaration of competing interest}
	The authors declare that they have no known competing financial interests or personal relationships that could have appeared to influence the work reported in this paper.
	
	\section*{Data availability}
	We have developed our research paper completely in an analytical approach. We did not produce any data for this publication. As a result, our research is not associated with any one form of data.
	
	\section*{Acknowledgments}
	We sincerely thank the referee/s for the critical comments and constructive suggestions, which have significantly improved the clarity, rigor, and overall quality of the manuscript. The authors S. H. Shekh  and Anirudh Pradhan  appreciate the help and resources provided by the IUCAA in Pune, India.
	
\end{document}